\documentclass[preprint,showpacs,preprintnumbers,amsmath,amssymb]{revtex4}

\usepackage{graphicx}
\usepackage{epsfig}		
\usepackage{dcolumn}
\usepackage{bm}

\def\0{\mbox{\tiny $0$}}
\def\1{\mbox{\tiny $1$}}
\def\2{\mbox{\tiny $2$}}
\def\3{\mbox{\tiny $3$}}
\def\4{\mbox{\tiny $4$}}
\def\5{\mbox{\tiny $5$}}
\def\6{\mbox{\tiny $6$}}
\def\7{\mbox{\tiny $7$}}
\def\8{\mbox{\tiny $8$}}
\def\9{\mbox{\tiny $9$}}

\def\n{\mbox{\tiny $n$}}
\def\k{\mbox{\tiny $k$}}
\def\kk{\mbox{\small $k$}}

\def\f14{\mbox{\tiny $\frac{1}{4}$}}

\def\L{\mbox{\tiny $L$}}
\def\R{\mbox{\tiny $R$}}
\def\B{\mbox{\tiny $B$}}
\def\ii{\mbox{\tiny $i$}}

\def\s{\mbox{\tiny $s$}}

\def\j{\mbox{\tiny $j$}}
\def\mi{\mbox{\tiny $-$}}
\def\ig{\mbox{\tiny $=$}}
\def\pl{\mbox{\tiny $+$}}
\def\ppm{\mbox{\tiny $\pm$}}

\def\bb#1{\mbox{\footnotesize $(#1)$}}

\begin{document}

\title{Theoretical correlation between possible evidences of neutrino chiral oscillations and polarization measurements}

\author{A. E. Bernardini}
\email{alexeb@ifi.unicamp.br}
\author{M. M. Guzzo}
\email{guzzo@ifi.unicamp.br}
\affiliation{Instituto de F\'{\i}sica Gleb Wataghin, UNICAMP,\\
PO Box 6165, 13083-970, Campinas, SP, Brasil.}

\date{\today}

\begin{abstract}
Reporting about the formalism with the Dirac equation we describe the dynamics of chiral oscillations for a fermionic particle non-minimally coupling with an external magnetic field.
For massive particles, the chirality and helicity quantum numbers represent different physical quantities of representative importance in the study of chiral interactions, in particular, in the context of neutrino physics.
After solving the interacting Hamiltonian (Dirac) equation for the corresponding {\em fermionic} Dirac-{\em type} particle (neutrino) and quantifying chiral oscillations in the Dirac wave packet framework, we avail the possibility of determining realistic neutrino chirality conversion rates by means of (helicity) polarization measurements. 
We notice that it can become feasible for some particular magnetic field configurations with large values of {\boldmath$B$} orthogonal to the direction of the propagating particle.
\end{abstract}

 

\maketitle

In spite of their electric charge neutrality, neutrinos can interact with photons through loop (radiative) diagrams which represent the interaction between a fermionic field $\psi\bb{x}$ and an electromagnetic field $A^{\mu}\bb{x}$.
The correspondent Lagrangian term is written in terms of the field-strength tensor $F^{\mu\nu}\bb{x}= \partial^{\mu}A^{\nu}\bb{x} - \partial^{\nu}A^{\mu}\bb{x}$ as
\small\begin{equation}
\mathcal{L} = \frac{1}{2}\,\overline{\psi}\bb{x} \,\sigma_{\mu\nu}\left[\mu\, F^{\mu\nu}\bb{x} - d \,\mathcal{F}^{\mu\nu}\bb{x}\right] \psi\bb{x} + h.c.
\end{equation}\normalsize
where $x = \bb{t, \mbox{\boldmath$x$}}$, $\sigma_{\mu\nu} = \frac{i}{2}[\gamma_{\mu},\gamma_{\nu}]$, the {\em dual} field-strength tensor $\mathcal{F}^{\mu\nu}\bb{x}$ is given by $\mathcal{F}^{\mu\nu}\bb{x}= \frac{1}{2}\epsilon^{\mu\nu\lambda\delta} F^{\lambda\delta}\bb{x}$\footnote{$\epsilon^{\mu\nu\lambda\delta}$ is the totally fourth rank antisymmetric tensor.} and the coefficients $\mu$ and $d$ respectively represent the magnetic and the electric dipole moment which establish the neutrino electromagnetic coupling.
Circumvented by this extensively explored scenario \cite{Vol81,Oli90,Ter90,Stu03,Stu04}, our aim is to partially accommodate the Dirac wave packet formalism \cite{Zub80,Ber04} with this further class of static characteristics of neutrinos, namely, the (electro)magnetic moment which appears in a Lagrangian with non-minimal coupling.
By following this line of reasoning, we evaluate the corresponding Hamiltonian equation, from which we can describe the dynamics of the chiral projection operator ($\gamma_{\5}$) and, consequently, determine the complete expression for the neutrino chirality conversion rate.
Thus we analyze the quantum oscillation phenomena for some particular magnetic field configurations with large values of {\boldmath$B$} parallel to the direction of the propagating particle.
We firstly set the non-minimally interacting (Dirac) Hamiltonian 
\small\begin{eqnarray} 
\mathit{H} &=& 
		    \mbox{\boldmath$\alpha$}\cdot \mbox{\boldmath$p$} + \beta \left[m - \mu\, \mbox{\boldmath$\Sigma$}\cdot \mbox{\boldmath$B$}\bb{x} + d \, \mbox{\boldmath$\Sigma$}\cdot \mbox{\boldmath$E$}\bb{x}\right] 
\label{01},~~
\end{eqnarray}\normalsize 
in terms of the Dirac matrices, $\mbox{\boldmath$\alpha$} = \sum_{\k \ig \1}^{\3} \alpha_{\k}\hat{\kk} = \sum_{\k \ig \1}^{\3} \gamma_{\0}\gamma_{\k}\hat{\kk}$,
$\beta = \gamma_{\0}$, where $\mbox{\boldmath$B$}\bb{x}$ and $\mbox{\boldmath$E$}\bb{x}$ are respectively the magnetic and electric fields.
In fact, the Eq.~(\ref{01}) could be extended to an equivalent matrix representation with flavor and mass mixing elements $(m_{\ii,\j})$ where the diagonal (off-diagonal) elements described by $\mu_{\ii,\j}$ and $d_{\ii,\j}$, where $i, \, j$ are mass indices, would be called diagonal (transition) moments.
In addition, for both Dirac and Majorana neutrinos, we could have transition amplitudes with non-vanishing magnetic and electric dipole moments \cite{Sch1,Mar1,Akh1}.
Otherwise, the CP invariance holds the diagonal electric dipole moments null \cite{Kim93}.
Particularly for Majorana neutrinos, it can be demonstrated that the diagonal magnetic and electric dipole moments vanish if $CPT$ invariance is assumed \cite{Sch1}.
Turning back to the simplifying example of diagonal moments and assuming CP and CPT invariance, we can restrict our analysis to the coupling with only an external magnetic field $\mbox{\boldmath$B$}\bb{x}$ by setting $d = 0$.
From this point, the expression for $\mu$ can be found from Feynman diagrams for magnetic moment corrections \cite{Kim93} and turns out to be proportional to the neutrino mass (matrix),
$\mu = \frac{3\, e \,G}{8 \sqrt{2}\pi^{\2}} m 
= 2.7 \times 10^{\mi \1\0}\,\mu_{\B}\,\frac{m_{\nu}}{m_N}$
where $G$ is the Fermi constant and $m_{N}$ is the nucleon mass\footnote{We are using some results of the standard $SU(2)_{L} \otimes U(1)_{Y}$ electroweak theory \cite{Gla61,Wei67,Sal68}.}.

Assuming the simplifying hypothesis of a uniform magnetic field  $\mbox{\boldmath$B$}$, the physical implications of the non-minimal coupling with an external magnetic field can then be studied by means of the eigenvalue problem expressed by the Hamiltonian equation
\small\begin{eqnarray} 
H\bb{\mbox{\boldmath$p$}} \, \varphi_{\n} = \,E_{\n}\bb{\mbox{\boldmath$p$}} \, \varphi_{\n}
 		   &=& \left\{\mbox{\boldmath$\alpha$}\cdot \mbox{\boldmath$p$} + \beta \left[m - \mu\, \mbox{\boldmath$\Sigma$}\cdot \mbox{\boldmath$B$}\right]\right\}\varphi_{\n}
\label{10}.
\end{eqnarray}\normalsize 
The most general eigenvalue ($E_{\n}\bb{\mbox{\boldmath$p$}}$) solution of the above problem is given by 
\small\begin{eqnarray} 
 E_{\n}\bb{\mbox{\boldmath$p$}} = \,  \pm E_{\s}\bb{\mbox{\boldmath$p$}} 
&=& \pm \sqrt{m^{\2} + \mbox{\boldmath$p$}^{\2} + \mbox{\boldmath$a$}^{\2} +\bb{\mi 1}^{\s}2
\sqrt{m^{\2}\mbox{\boldmath$a$}^{\2} + \bb{\mbox{\boldmath$p$} \times \mbox{\boldmath$a$}}^{\2}}}, ~~~~s\,=\, 1,\,2
\label{11},
\end{eqnarray}\normalsize 
where we have denoted $E_{\n \ig \1,\2,\3,\4} = \pm E_{\s \ig \1,\2}$, and we have set $\mbox{\boldmath$a$} = \mu\, \mbox{\boldmath$B$}$.
The complete set of orthonormal eigenstates $\varphi_{\n}$ thus can be written in terms of the eigenfunctions $\mathcal{U}\bb{p_{\s}}$ with positive energy eigenvalues ($+ E_{\s}\bb{\mbox{\boldmath$p$}}$) and the eigenfunctions $\mathcal{V}\bb{p_{\s}}$ with negative energy eigenvalues ($- E_{\s}\bb{\mbox{\boldmath$p$}}$),
\small\begin{eqnarray} 
\mathcal{U}\bb{p_{\s}} &=& -N\bb{p_{\s}}\, \left\{\sqrt{\frac{A^{\mi}_{\s}}{A^{\pl}_{\s}}},\,\sqrt{\frac{\alpha^{\pl}_{\s}}{\alpha^{\mi}_{\s}}},\,\sqrt{\frac{A^{\mi}_{\s}\alpha^{\pl}_{\s}}{A^{\pl}_{\s}\alpha^{\mi}_{\s}}},\,-1\right\}^{\dagger}\nonumber\\
\mathcal{V}\bb{p_{\s}} &=& -N\bb{p_{\s}}\, \left\{\sqrt{\frac{A^{\mi}_{\s}}{A^{\pl}_{\s}}},\,-\sqrt{\frac{\alpha^{\mi}_{\s}}{\alpha^{\pl}_{\s}}},\,-\sqrt{\frac{A^{\mi}_{\s}\alpha^{\mi}_{\s}}{A^{\pl}_{\s}\alpha^{\pl}_{\s}}},\,-1\right\}^{\dagger}
\label{12},
\end{eqnarray}\normalsize
where $p_{\s}$ is the relativistic {\em quadrimomentum}, $p_{\s} = (E_{\s}\bb{\mbox{\boldmath$p$}}, \mbox{\boldmath$p$})$, $N\bb{p_{\s}}$ is the normalization constant and $A^{\ppm}_{\s} = \Delta_{\s}^{\2}\bb{\mbox{\boldmath$p$}} \pm 2 m |\mbox{\boldmath$a$}|  - \mbox{\boldmath$a$}^{\2}$, $\alpha^{\ppm}_{\s}= 2 E_{\s}\bb{\mbox{\boldmath$p$}} |\mbox{\boldmath$a$}| \pm (\Delta_{\s}^{\2}\bb{\mbox{\boldmath$p$}} + \mbox{\boldmath$a$}^{\2})$ with $\Delta_{\s}^{\2}\bb{\mbox{\boldmath$p$}} = E_{\s}^{\2}\bb{\mbox{\boldmath$p$}} - (m^{\2} + \mbox{\boldmath$p$}^{\2}) + \mbox{\boldmath$a$}^{\2}$.
We can observe that the above spinorial solutions are free of any additional constraint, namely, at a given time $t$, they are independent functions of $\mbox{\boldmath$p$}$ and they do not represent chirality/helicity eigenstates \cite{Kha05,Kha01,Roc06}.
The time-evolution of a plane-wave packet $\psi\bb{t, \mbox{\boldmath$x$}}$ can be written as
\small\begin{eqnarray}
\psi\bb{t, \mbox{\boldmath$x$}}
&=& \int\hspace{-0.1 cm} \frac{d^{\3}\hspace{-0.1cm}\mbox{\boldmath$p$}}{(2\pi)^{\3}}
\sum_{\s \ig \1,\2}\{b\bb{p_{\s}}\mathcal{U}\bb{p_{\s}}\, \exp{[- i\,E_{\s}\bb{\mbox{\boldmath$p$}}\,t]}
+ d^*\bb{\tilde{p}_{\s}}\mathcal{V}\bb{\tilde{p}_{\s}}\, \exp{[+i\,E_{\s}\bb{\mbox{\boldmath$p$}}\,t]}\}
\exp{[i \, \mbox{\boldmath$p$} \cdot \mbox{\boldmath$x$}]},
\label{14}
\end{eqnarray}\normalsize
with $\tilde{p}_{\s} = (E_{\s},-\mbox{\boldmath$p$})$.
Meanwhile, the Eq.~(\ref{14}) requires some extensive mathematical manipulations for explicitly constructing the dynamics of a generic operator $\mathcal{O}\bb{t}$.
If, however, the quoted observables like the chirality $\gamma^{\5}$, the helicity $h$ or even the spin projection onto $\mbox{\boldmath$B$}$ commuted with the Hamiltonian $H$, we could reconfigure the above solutions to simpler ones. 
Let us then assume that the magnetic field $\mbox{\boldmath$B$}$ is either orthogonal or parallel to the momentum $\mbox{\boldmath$p$}$.
For both of these cases the spinor eigenstates can then be decomposed into orthonormal bi-spinors as $\varphi^{\ppm}_{\1,\2}$ and $\chi^{\ppm}_{\1,\2}$ which are eigenstates of the spin projection operator $\mbox{\boldmath$\sigma$}\cdot\mbox{\boldmath$B$}$, i. e. beside of being energy eigenstates, the general solutions $\mathcal{U}\bb{p_{\s}}$ and $\mathcal{V}\bb{p_{\s}}$ would become eigenstates of the operator $\mbox{\boldmath$\Sigma$}\cdot\mbox{\boldmath$B$}$ and, equivalently, of $\mbox{\boldmath$\Sigma$}\cdot\mbox{\boldmath$a$}$.
Now the Eq.(\ref{10}) can be decomposed into a pair of coupled equations like
\small\begin{eqnarray}
\left(\pm E_{\s} - m + \mbox{\boldmath$\sigma$}\cdot\mbox{\boldmath$a$} \right)\varphi^{\ppm}_{\s} &=& \pm \mbox{\boldmath$\sigma$}\cdot\mbox{\boldmath$p$}\,\chi^{\ppm}_{\s},\nonumber\\
\left(\pm E_{\s} + m - \mbox{\boldmath$\sigma$}\cdot\mbox{\boldmath$a$} \right)\chi^{\ppm}_{\s} &=& \pm \mbox{\boldmath$\sigma$}\cdot\mbox{\boldmath$p$}\,\varphi^{\ppm}_{\s},
\label{17}
\end{eqnarray}\normalsize
where we have suppressed the $p_{\s}$ dependence.

By introducing the anti-commuting relation $\{\mbox{\boldmath$\sigma$}\cdot\mbox{\boldmath$p$},\,\mbox{\boldmath$\sigma$}\cdot\mbox{\boldmath$B$}\} = 0$ derived when $\mbox{\boldmath$p$}\cdot\mbox{\boldmath$B$} = 0$, the eigenspinor representation can be reduced to
\small\begin{equation}
\mathcal{U}\bb{p_{\s}} = \sqrt{\frac{\varepsilon_{\0} + m}{2\varepsilon_{\0}}}
\left[\begin{array}{r} \varphi^{\pl}_{\s}\\ \frac{\mbox{\boldmath$\sigma$}\cdot\mbox{\boldmath$p$}}{\varepsilon_{\0}+ m}\,\varphi^{\pl}_{\s}\end{array}\right]
~~\mbox{and}~~
\mathcal{V}\bb{p_{\s}} = \sqrt{\frac{\varepsilon_{\0} + m}{2\varepsilon_{\0}}}
\left[\begin{array}{r} \frac{\mbox{\boldmath$\sigma$}\cdot\mbox{\boldmath$p$}}{\varepsilon_{\0}+ m}\,\chi^{\mi}_{\s}\\ \chi^{\mi}_{\s}\end{array}\right]
\label{21},
\end{equation}\normalsize
with $\varepsilon_{\0} = \sqrt{\mbox{\boldmath$p$}^{\2} + m^{\2}}$ and the energy eigenvalues $\pm E_{\s} = \pm\left[\varepsilon_{\0} + \bb{\mi 1}^{\s}|\mbox{\boldmath$a$}|\right]$.

Analogously, by introducing the commuting relation $[\mbox{\boldmath$\sigma$}\cdot\mbox{\boldmath$p$},\,\mbox{\boldmath$\sigma$}\cdot\mbox{\boldmath$B$}] = 0$ when $\mbox{\boldmath$p$}\times\mbox{\boldmath$B$} = 0$, the eigenspinor representation can be reduced to
\small\begin{equation}
\mathcal{U}\bb{p_{\s}} = \sqrt{\frac{E_{\s} + m_{\s}}{2E_{\s}}}
\left[\begin{array}{r} \varphi^{\pl}_{\s}\\ \frac{\mbox{\boldmath$\sigma$}\cdot\mbox{\boldmath$p$}}{E_{\s}+ m_{\s}}\,\varphi^{\pl}_{\s}\end{array}\right]
~~\mbox{and}~~
\mathcal{V}\bb{p_{\s}} = \sqrt{\frac{E_{\s} + m_{\s}}{2E_{\s}}}
\left[\begin{array}{r} \frac{\mbox{\boldmath$\sigma$}\cdot\mbox{\boldmath$p$}}{E{\s}+ m_{\s}}\,\chi^{\mi}_{\s}\\ \chi^{\mi}_{\s}\end{array}\right]
\label{25},
\end{equation}\normalsize
with $m_{\s} = m - \bb{\mi 1}^{\s}|\mbox{\boldmath$a$}|$ and the energy eigenvalues $\pm E_{\s} = \pm \sqrt{\mbox{\boldmath$p$}^{\2} + m_{\s}^{\2}}$.
Since we can set $\varphi^{\pl}_{\1,\2} \equiv \chi^{\mi}_{\1,\2}$ for both of the above cases as the components of an orthonormal basis, the orthogonality and the closure relations are immediate.

Finally, the calculation of the expectation value of $\gamma_{\5}\bb{t}$ is substantially simplified when we substitute the above closure relations into the wave-packet expression of Eq.~(\ref{14}).
To clarify this point, we suppose the initial condition over $\psi\bb{t,\mbox{\boldmath$x$}}$ can be set in terms of the Fourier transform of the weight function 
$\varphi\bb{\mbox{\boldmath$p$}-\mbox{\boldmath$p$}_{\0}}\,w \,=\,
\sum_{\s \ig \1,\2}{\{b\bb{p_{\s}}\mathcal{U}\bb{p_{\s}} + d^*\bb{\tilde{p}{\s}}\mathcal{V}\bb{\tilde{p}_{\s}}\}}$,
\small\begin{equation}
\psi\bb{0, \mbox{\boldmath$x$}}
= \int\hspace{-0.1 cm} \frac{d^{\3}\hspace{-0.1cm}\mbox{\boldmath$p$}}{(2\pi)^{\3}}
\varphi\bb{\mbox{\boldmath$p$}-\mbox{\boldmath$p$}_{\0}}\exp{[i \, \mbox{\boldmath$p$} \cdot \mbox{\boldmath$x$}]}
\,w
\label{29}
\end{equation}\normalsize
where $w$ is some fixed normalized spinor, and we have found \cite{Zub80} $b\bb{p_{\s}} = \varphi\bb{\mbox{\boldmath$p$}- \mbox{\boldmath$p$}_{\0}} \, \mathcal{U}^{\dagger}\bb{p_{\s}} \, w$ and $d^*\bb{\tilde{p}_{\s}} = \varphi\bb{\mbox{\boldmath$p$}- \mbox{\boldmath$p$}_{\0}}\,\mathcal{V}^{\dagger}\bb{\tilde{p}_{\s}}\, w$.
For {\em any} initial state $\psi\bb{0, \mbox{\boldmath$x$}}$ given by Eq.~(\ref{29}), the negative frequency solution coefficient $d^*\bb{\tilde{p}_{\s}}$ necessarily provides a non-null contribution to the time-evolving wave-packet.
This obliges us to take the complete set of Dirac equation solutions to construct a complete and correct wave-packet solution.
Only if we consider the initial spinor $w$ being a positive energy ($E_{\s}\bb{\mbox{\boldmath$p$}}$) and momentum $\mbox{\boldmath$p$}$ eigenstate, the contribution due to the negative frequency solutions $d^*\bb{\tilde{p}_{\s}}$ will become null and we will have a simple expression for the time-evolution of any physical observable.
By substituting the explicit form of the spinors of Eqs.~(\ref{21}) and (\ref{25}) (closure relations) into the time-evolution expression for the above wave-packet, the Eq.(\ref{14}) can thus be rewritten as
{\small \small\begin{eqnarray}
\hspace{-0.1 cm}\psi\bb{t, \mbox{\boldmath$x$}}
&\hspace{-0.1 cm}=&
\hspace{-0.1 cm}\int\hspace{-0.1 cm} \frac{d^{\3}\hspace{-0.1cm}\mbox{\boldmath$p$}}{(2\pi)^{\3}}
\varphi\bb{\mbox{\boldmath$p$}-\mbox{\boldmath$p$}_{\0}}\exp{[i \, \mbox{\boldmath$p$} \cdot \mbox{\boldmath$x$}]}
\sum_{\s \ig \1,\2}\mbox{$\left\{\left[\cos{[E_{\s}\,t]} -i\frac{H_{\0}}{\varepsilon_{\0}}\sin{[E_{\s}\,t]}\right]\left(\frac{1-(\mi 1)^{\s}\gamma_{\0}\mbox{\boldmath$\Sigma$}\cdot\hat{\mbox{\boldmath$a$}}}{2}\right)\right\}$}
w~~~~
\label{14B}
\end{eqnarray}\normalsize}
for the first case where $E_{\s} = \left[\varepsilon_{\0} + \bb{\mi 1}^{\s}|\mbox{\boldmath$a$}|\right]$ and $H_{\0} = \mbox{\boldmath$\alpha$}\cdot \mbox{\boldmath$p$} + \gamma_{\0} m$, or as
{\small \small\begin{eqnarray}
\hspace{-0.1 cm}\psi\bb{t, \mbox{\boldmath$x$}}
&\hspace{-0.1 cm}=&
\hspace{-0.1 cm}\int\hspace{-0.1 cm} \frac{d^{\3}\hspace{-0.1cm}\mbox{\boldmath$p$}}{(2\pi)^{\3}}
\varphi\bb{\mbox{\boldmath$p$}-\mbox{\boldmath$p$}_{\0}}\exp{[i \, \mbox{\boldmath$p$} \cdot \mbox{\boldmath$x$}]}
\sum_{\s \ig \1,\2}\mbox{$\left\{\left[\cos{[E_{\s}\,t]} -i\frac{H_{\s}}{E_{\s}}\sin{[E_{\s}\,t]}\right]\left(\frac{1-(\mi 1)^{\s}\mbox{\boldmath$\Sigma$}\cdot\hat{\mbox{\boldmath$a$}}}{2}\right)\right\}$}
w~~
\label{14A}
\end{eqnarray}\normalsize}
for the second case where $E_{\s} = \sqrt{\mbox{\boldmath$p$}^{\2} + m_{\s}^{\2}}$ and $H_{\s} = \mbox{\boldmath$\alpha$}\cdot \mbox{\boldmath$p$} + \gamma_{\0} m_{\s}$.

It was already demonstrated that, in vacuum, chiral oscillations can introduce very small modifications to the neutrino conversion formula \cite{Ber05,Ber06A}.
The probability of a neutrino produced as a negative chiral eigenstate be detected after a time $t$ can be summarized by the expression
\small\begin{eqnarray}
P(\mbox{\boldmath$\nu_{\alpha,\L}$}\rightarrow\mbox{\boldmath$\nu_{\alpha,\L}$};t) 
& = &
\int{d^{\3}\mbox{\boldmath$x$}\,\psi^{\dagger}\bb{t, \mbox{\boldmath$x$}}\,\frac{1 - \gamma_{\5}}{2}\,
\psi\bb{t, \mbox{\boldmath$x$}}} = \frac{1}{2}\left(1 - \langle\gamma_{\5}\rangle\bb{t}\right)
\label{15A}.
\end{eqnarray}\normalsize
From this integral, it is readily seen that an initial $\mi 1$ chiral mass-eigenstate will evolve with time changing its chirality.
By assuming the fermionic particle is created  at time $t=0$ as a $\mi 1$ chiral eigenstate ($\gamma_{\5} w = \mi w$), in case of $\{\mbox{\boldmath$\sigma$}\cdot\mbox{\boldmath$p$},\,\mbox{\boldmath$\sigma$}\cdot\mbox{\boldmath$B$}\} = 0$ ({\boldmath$B$} orthogonal to {\boldmath$p$}), we could write
\small\begin{eqnarray}
 \langle\gamma_{\5}\rangle\bb{t} &=&  
(\mi 1) \int\hspace{-0.1 cm} \frac{d^{\3}\hspace{-0.1cm}\mbox{\boldmath$p$}}{(2\pi)^{\3}}\varphi^{\2}\bb{\mbox{\boldmath$p$}-\mbox{\boldmath$p$}_{\0}}
\mbox{$\left\{\frac{\mbox{\boldmath$p$}^{\2}}{\epsilon_{\0}^{\2}}\cos{[2\,|\mbox{\boldmath$a$}|\,t]}+ \frac{ m^{\2}}{\epsilon_{\0}^{\2}} \cos{[2\,\epsilon_{\0}\,t]}\right\}$}
\label{15C},
\end{eqnarray}\normalsize
where we have used the wave packet expression of Eq.~(\ref{14B}) and, in addition to $w^{\dagger} \gamma_{\5} w = \mi 1$, we have also observed that $\{H_{\0},\,\gamma_{\5}\} = 2 \gamma_{\5}\mbox{\boldmath$\Sigma$}\cdot\hat{\mbox{\boldmath$p$}}$ and, subsequently, $w^{\dagger}\mbox{\boldmath$\Sigma$}\cdot\hat{\mbox{\boldmath$p$}}\gamma_{\0}\mbox{\boldmath$\Sigma$}\cdot\hat{\mbox{\boldmath$a$}} w = 0$.
For the above result, in addition to the non-interacting oscillating term $\frac{ m^{\2}}{\epsilon_{\0}^{\2}} \cos{[2\,\epsilon_{\0}\,t]}$, which comes from the interference between positive and negative frequency solutions of the Dirac equation, we have an extra term which comes from the interference between equal sign frequencies and, for very large time scales, can substantially change the oscillating results. 
In this case, it is interesting to observe that the ultra-relativistic limit of Eq.(\ref{15C}) leads to the following expressions for the chiral conversion formulas,
\small\begin{equation}
P(\mbox{\boldmath$\nu_{\alpha,\L}$}\rightarrow\mbox{\boldmath$\nu_{\alpha,\L(\R)}$};t) 
 \approx  \frac{1}{2}\left(1 +(-) \cos{[2\,|\mbox{\boldmath$a$}|\,t]}\right)
\label{15A2}
\end{equation}\normalsize
which, differently from chiral oscillations in vacuum, can be phenomenologically relevant.
Obviously, we are reproducing the consolidated results already attributed to neutrino spin-flipping \cite{Kim93} where, by taking the ultra-relativistic limit, the chirality quantum number can be approximated by the helicity quantum number, but now it was accurately derived from the complete formalism with Dirac spinors.
We still remark that, in the standard treatment of vacuum neutrino oscillations, the use of scalar mass-eigenstate wave packets made up exclusively of positive frequency plane-wave solutions is implicitly assumed. 

Turning back to the case where $[\mbox{\boldmath$\sigma$}\cdot\mbox{\boldmath$p$},\,\mbox{\boldmath$\sigma$}\cdot\mbox{\boldmath$B$}] = 0$ ({\boldmath$B$} parallel to {\boldmath$p$}), we could have a phenomenologically more interesting result. 
By following a similar procedure with the mathematical manipulations, we could write
{\small \begin{eqnarray}
 \langle\gamma_{\5}\rangle\bb{t}&=&  
(\mi 1) \int\hspace{-0.1 cm} \frac{d^{\3}\hspace{-0.1cm}\mbox{\boldmath$p$}}{(2\pi)^{\3}}\varphi^{\2}\bb{\mbox{\boldmath$p$}-\mbox{\boldmath$p$}_{\0}}
\sum_{\s \ig \1,\2}\mbox{$\left\{\left[\frac{\mbox{\boldmath$p$}^{\2}}{E_{\s}^{\2}} + \frac{m_{\s}^{\2}}{E_{\s}^{\2}}\cos{[2\,E_{\s}\,t]}\right]\,w^{\dagger}\left(\frac{1-(\mi 1)^{\s}\mbox{\boldmath$\Sigma$}\cdot\hat{\mbox{\boldmath$a$}}}{2}\right)w\right\}$}
\label{15B},
\end{eqnarray}\normalsize}
where we have used the wave packet expression of Eq.~(\ref{14A}) and, in the second passage, we have observed that $w^{\dagger} \gamma_{\5} w = \mi 1$, $w^{\dagger} [H_{\s},\,\gamma_{\5}]w = 0$ and $H_{\s}\,\gamma_{\5}\,H_{\s} = \mbox{\boldmath$p$}^{\2} - m_{\s}^{\2}$.
The above expression can be reduced to a simpler one in the non-interacting case where just the mass term appears.
Due to a residual interaction with the external magnetic field {\boldmath$B$}, we could also observe chiral oscillations in the ultra-relativistic limit.
As a {\em toy model} illustration, by assuming a highly peaked momentum distribution centered around a non-relativistic momentum $p_{\ii}\ll m_{\s}$, where the wave packet effects are practically ignored, the chiral conversion formula can be written as
\small\begin{equation}
 P(\mbox{\boldmath$\nu_{\alpha,\L}$}\rightarrow\mbox{\boldmath$\nu_{\alpha,\R}$};t) 
 \approx  \frac{1}{2}\left(1 - \cos{[2\,m\,t]}\cos{[2\,|\mbox{\boldmath$a$}|\,t]}
 -\sin {[2\,m\,t]}\sin{[2\,|\mbox{\boldmath$a$}|\,t]}w^{\dagger}\mbox{\boldmath$\Sigma$}\cdot\hat{\mbox{\boldmath$a$}}w\right)
\label{15A222}
\end{equation}\normalsize
where all the oscillating terms come from the interference between positive and negative frequency solutions which compose the wave packets.

\begin{figure}
\vspace{-0.7cm}
\begin{center}
\epsfig{file= 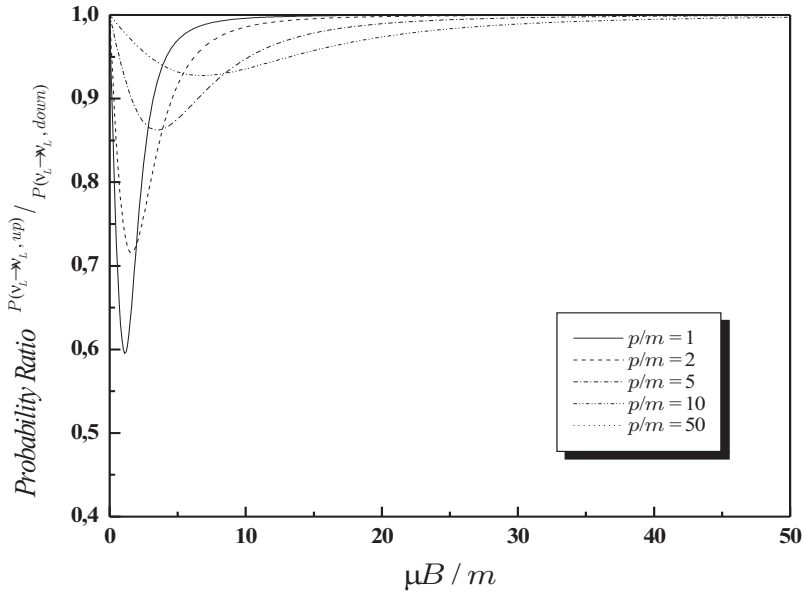, height= 10 cm, width= 12 cm}
\end{center}
\vspace{-1.7cm}
\small\caption{\label{Chi1} We plot the ratio between the detection probabilities of spin-up and spin-down negative chiral eigenstates as a function of the magnetic interaction energy $\mu\mbox{\boldmath{$B$}}/m$ for different values of $\mbox{\boldmath{$p$}}_{\0}/m$ representing the propagation regime.
We have established that $w^{\dagger}\left(\frac{1-(\mi 1)^{\1}\mbox{\boldmath$\Sigma$}\cdot\hat{\mbox{\boldmath$a$}}}{2}\right)w = w^{\dagger}\left(\frac{1-(\mi 1)^{\2}\mbox{\boldmath$\Sigma$}\cdot\hat{\mbox{\boldmath$a$}}}{2}\right)w = \frac{1}{2}$ which is equivalent to the assumption of neutrinos being created at $t=0$ with a completely random spin orientation, which can be viewed as a collection of neutrinos where one-half are characterized by spin-up states and the remaining one-half by spin-down states.}
\end{figure}

\begin{figure}
\vspace{-0.7cm}
\begin{center}
\epsfig{file= 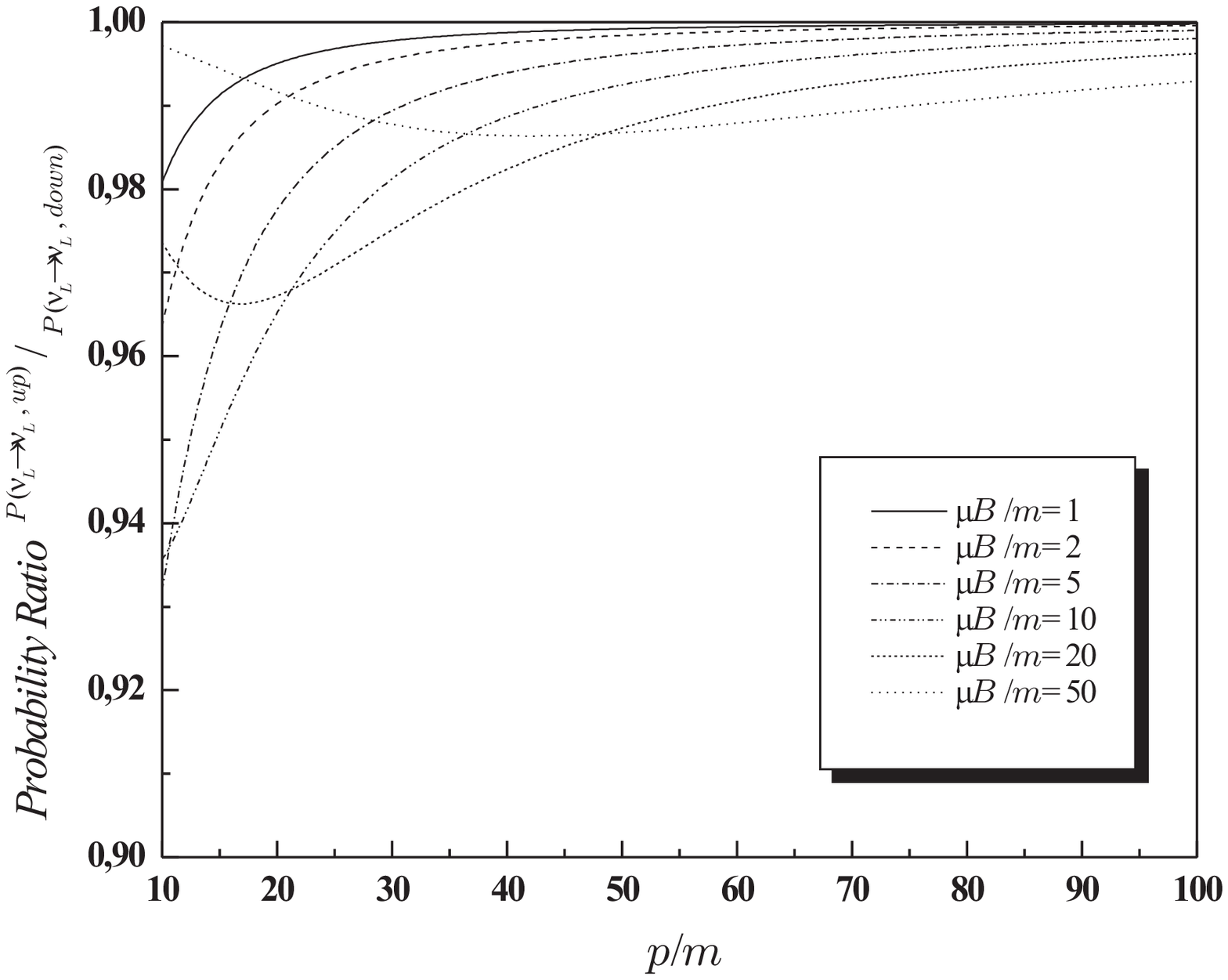, height= 10 cm, width= 12 cm}
\end{center}
\vspace{-1.7cm}
\small\caption{\label{Chi2}  We plot the ratio between the detection probabilities of spin-up and spin-down negative chiral eigenstates as a function of $\mbox{\boldmath{$p$}}_{\0}/m$ representing the propagation regime for different values of the magnetic interaction energy $\mu\mbox{\boldmath{$B$}}/m$.
We have also established that $w^{\dagger}\left(\frac{1-(\mi 1)^{\1}\mbox{\boldmath$\Sigma$}\cdot\hat{\mbox{\boldmath$a$}}}{2}\right)w = w^{\dagger}\left(\frac{1-(\mi 1)^{\2}\mbox{\boldmath$\Sigma$}\cdot\hat{\mbox{\boldmath$a$}}}{2}\right)w = \frac{1}{2}$.}
\end{figure}

By assuming that the neutrinos before interacting with the magnetic fields are produced with completely random spin orientation, which can be viewed as a collection of neutrinos where $50\%$ are characterized by spin-up states and the remaining $50\%$ by spin-down states, as we have noted before, we can understand the result provided by the Eq.~(\ref{15B}) as a phenomenological instrument much more interesting than the result expressed in the Eq.~(\ref{15C}).
By averaging on time both the above results obtained for $\langle\gamma_{\5}\rangle\bb{t}$, we can easily observe that the corresponding average value for the case where {\boldmath$p$} is perpendicular to {\boldmath$B$} is zero, and the same average value for the case where {\boldmath$p$} is parallel to {\boldmath$B$} depends on the polarization of the initial state as we can observe in the following expression,
{\small\begin{eqnarray}
\langle \langle\gamma_{\5}\rangle\bb{t}\rangle_{time}&=&  
(\mi 1) \int\hspace{-0.1 cm} \frac{d^{\3}\hspace{-0.1cm}\mbox{\boldmath$p$}}{(2\pi)^{\3}}\varphi^{\2}\bb{\mbox{\boldmath$p$}-\mbox{\boldmath$p$}_{\0}}
\sum_{\s \ig \1,\2}\mbox{$\left[\frac{\mbox{\boldmath$p$}^{\2}}{E_{\s}^{\2}}
\,w^{\dagger}\left(\frac{1-(\mi 1)^{\s}\mbox{\boldmath$\Sigma$}\cdot\hat{\mbox{\boldmath$a$}}}{2}\right)
w\right]$}
\label{15BB},
\end{eqnarray}\normalsize}
By supposing that the construction of a detection apparatus which identify the neutrino state of polarization (helicity!) is feasible, the last result becomes phenomenologically relevant since it distinguish the chirality conversion rates for spin-up ($s=1$) and spin-down ($s=2$) states.
Once we have assumed the neutrino electroweak interactions at the source and detector are ({\em left} or negative) chiral $\left(\overline{\psi} \gamma^{\mu}(1 - \gamma^{\5})\psi W_{\mu}\right)$ only the component with negative chirality contributes to the effective detection result.
The {\em residual} contribution due to the chiral oscillation mechanism appears in the final time averaged result. 
It leads to different expected probabilities of detecting a negative chiral eigenstate for the opposite polarization states.
By comparing the initial polarization states where, before interacting with the magnetic fields, for any detection process, the spin-up and spin-down states would be equally probable, with the expected results for the polarization states after traversing the magnetized region with {\boldmath$p$} parallel to {\boldmath$B$}, where the chiral oscillation mechanism effectively takes place, we observe that the chiral oscillation mechanism can lead to a modification on the final expectation values.
It reflects in a ratio between the spin-up and spin-down negative chiral eigenstate probabilities different from the unity as we have illustrated in the Figs.~\ref{Chi1} and \ref{Chi2}.

As one can notice in the plots, since the coefficient of the oscillating term goes with $\frac{m_{\s}^{\2}}{E_{\s}^{\2}}$, the chiral oscillation effects are not relevant for ultra-relativistic neutrinos \cite{Ber05,Ber05A}.
At the same time, there is a natural scale for the magnetic field strength that is required to have a significant impact on the quantum process, i. e. a critical value of $4.41 \times 10^{\1\3}\, G$.
There are reasons to expect that magnetic fields of such or even larger magnitudes can arise in cataclysmic astrophysical events of such as supernova explosions or coalescing neutron stars, situations where an enormous neutrino outflow should be expected.
Two classes of stars, the soft gamma-ray repeaters (SGR) \cite{SGR} and the anomalous x-ray pulsars (AXP) \cite{AXP,AXP2} are supposed to be remnants of such events which form magnetars \cite{MAG}, neutron stars with magnetic fields of $10^{\1\4} - 10^{\1\5} \, G$.
From the theoretical point of view, the possibility of larger magnetic fields $10^{\1\6} - 10^{\1\7} \, G$ have not been discarded yet.
The early universe between the QCD phase transition ($\sim 10^{-\5} \, s$) and the nucleosynthesis epoch ($10^{-\2} - 10^{\2} \, s$) is believed to be yet another natural environment where strong magnetic fields and large neutrino densities could coexist.

In which concern with the polarization measurements, the observation of the dependence of the neutrino chirality conversion rate on the neutrino polarization state (measurements) can be converted into a clear signal of the presence of right-handed (positive chiral) neutrinos in the neutrino-electron scattering.
In fact, in a more extended scenario, the scattering of neutrinos on a polarized electron target \cite{Mis05} was proposed as a test for new physics beyond the Standard Model (SM).
To search for exotic right-handed weak interactions \cite{Mis05,Cie05} the strong polarized neutrino beam and the polarized neutrino target is required \cite{Mis05,Cie05}.
It has been shown how the presence of the right-handed neutrinos changes the spectrum of recoil electrons in relation to the expected standard model prediction, using the current limits on the non-standard couplings.
In this framework, the interference terms between the standard and exotic couplings in the differential cross section depend on the angle between the transverse incoming neutrino polarization and the transverse electron polarization of the target, and do not vanish in the limit of massless neutrino.

Finally, we would have been dishonest if we had ignored the complete analysis of the general case comprised by Eqs.~(\ref{10}-\ref{12}) where we had not yet assumed an arbitrary (simplified) spatial configuration for the magnetic field.
We curiously notice the fact that those complete (general) expressions for propagating wave packets do not satisfy the standard dispersion relations like $E^{\2} = m^{\2}+{\mbox{\boldmath$p$}}^{\2}$ excepting by the two particular cases where $E_{\s}\bb{\mbox{\boldmath$p$}}^{\2} = m_{\s}^{\2}+ {\mbox{\boldmath$p$}}^{\2}$ for $\mbox{\boldmath$p$}\times\mbox{\boldmath$B$} = 0$ or $\epsilon_{\0}^{\2} = m^{\2}+ {\mbox{\boldmath$p$}}^{\2}$ for $\mbox{\boldmath$p$}\cdot\mbox{\boldmath$B$} = 0$.
In particular, the process of neutrino propagation through an active medium consisting of magnetic field and plasma and the consequent modifications to the neutrino dispersion relations have been studied in the literature \cite{Kuz06}. 
In addition, such a general case leads to the formal connection between quantum oscillation phenomena and a very different field.
By principle, it could represent an inconvenient obstacle forbidding  the extension of these restrictive cases to a general one.
However, we believe that it can also represent a starting point in discussing dispersion relations which may be incorporated into frameworks encoding the breakdown (or the validity) of Lorentz invariance. 

Turning back to the foundations for applying the intermediate wave packet formalism in the neutrino oscillation problem, we know the necessity of a more sophisticated approach is required.
It can involve a field-theoretical treatment.
Derivations of the oscillation formula resorting to field-theoretical methods are not very popular.
They are thought to be very complicated and the existing quantum field computations of the oscillation formula do not agree in all respects \cite{Beu03}.
The  Blasone and Vitiello (BV) model \cite{Bla95,Bla03} to neutrino/particle mixing and oscillations seems to be the most distinguished trying to this aim.
They have attempt to define a Fock space of weak eigenstates to derive a nonperturbative oscillation formula.
Also with Dirac wave-packets, the flavor conversion formula can be reproduced \cite{Ber04} with the same mathematical structure as those obtained in the BV model \cite{Bla95,Bla03}.

To summarize, in the context where we have intended to explore the Dirac formalism, we have assumed that a satisfactory description for understanding chiral oscillations of fermionic (spin one-half) particles like neutrinos requires the use of the Dirac equation as evolution equation for the mass-eigenstates.
With such a framework stated, we have added the non-minimal coupling of the neutrino magnetic moment with an electromagnetic field in an external interacting process \cite{Vol81}.
Although clear experimental evidences are still missing, we have reinforced the idea that the spinorial form and the interference between positive and negative frequency components of the mass-eigenstate wave packets can introduce small modifications to the {\em standard} conversion formulas where chiral oscillations are taken into account, in particular, under circumstances where we treat the phenomenological possibility of neutrino interactions with very large magnetic fields.
For the future outline, we intend to inspect the correlated constructions with more general magnetic field spatial configurations as well as we plan to aggregate the corrections discussed here to a more generalized study of chiral conversion rates, polarization effects and neutrino propagation in magnetized media which have been already under consideration in the literature, in particular, those which could be interesting from the point of view of observable effects or phenomenological implications.
Also in the theoretical framework, we can notice a more extended discussion involving the evolution equation of the covariant neutrino spin operator in the Heisenberg representation in presence of general external fields \cite{Stu04}.
Such an analysis involves both electromagnetic and weak interactions with background matter \cite{Stu03,Stu04}.
In a similar context, some developments considering the minimal coupling of an electrically charged particle with an external magnetic field were also performed for an electron described by the Dirac equation \cite{Ter90} and the theory of spin light of neutrino in matter and electromagnetic fields has been extensively studied \cite{Stu04}.

{\bf Acknowledgments}
The authors would like to thank FAPESP and CNPq for the financial support of this work.

\end{document}